\documentclass[DIV=14, letterpaper]{scrartcl}

\usepackage{url}

\usepackage{amsmath,amssymb,amsfonts,subfig} 
\usepackage{graphicx}

\begin{document}

\date{}
\title{A Preliminary Comparison Between Compressive Sampling and Anisotropic Mesh-based Image Representation}
\author{Xianping Li
\thanks{
College of Integrative Sciences and Arts, Arizona State University, Mesa, AZ 85212, 
U.S.A. (\textit{Xianping.Li@asu.edu})}
\and Teresa Wu 
\thanks{
School of Computing, Informatics, Decision Systems Engineering, Arizona State University, Tempe, AZ 85281,
U.S.A. (\textit{Teresa.Wu@asu.edu})}
}

\maketitle

\textbf{Abstract}
\begin{abstract}
Compressed sensing (CS) has become a popular field in the last two decades to represent and reconstruct a sparse signal with much fewer samples than the signal itself. Although regular images are not sparse on their own, many can be sparsely represented in wavelet transform domain. Therefore, CS has also been widely applied to represent digital images. However, an alternative approach, adaptive sampling such as mesh-based image representation (MbIR), has not attracted as much attention. MbIR works directly on image pixels and represents the image with fewer points using a triangular mesh. In this paper, we perform a preliminary comparison between the CS and a recently developed MbIR method, AMA representation. The results demonstrate that, at the same sample density, AMA representation can provide better reconstruction quality than CS based on the tested algorithms. Further investigation with recent algorithms is needed to perform a thorough comparison. \\
\end{abstract}

\noindent
\textbf{Keywords.} {Compressive sampling, mesh-based image representation, AMA representation, PSNR, Structural similarity index measure}

\section{Introduction}
\label{sec-intro}

Compressive sampling (CS, also known as compressive sensing, compressed sensing, sparse sampling) has become popular in signal processing for the last two decades, and numerous research has been performed in this field. The literature is superabundant to be listed entirely, and only a few are referred here \cite{CRT06, Don06, Cha07, DT08, BD08, DM09, NV10, KSLB16}. CS has been applied to computed tomography (CT) \cite{TJYPJ11, XS11} and MRI \cite{Lus07, LDSP08, TA18} reconstructions to improve the acquisition speed. It has also been applied to represent digital images obtained from different sources \cite{FWX13, LXF17}, although many images are not sparse on their own.

In this paper, we focus on the representation of digital images from different sources, such as digital cameras, remote sensing images, as well as MR images that have already been obtained/reconstructed. In many processing tasks of those images, it is also desirable to represent the images with fewer pixels to reduce the image size and improve processing efficiency. One example is image segmentation via a variational approach. In those cases, we need to work on the image pixels directly.

Adaptive sampling based on image pixels, especially, triangular mesh-based image representation (MbIR), has recently gained much interest, although not as much as compressive sampling that works on transformed domain. Some publications about MbIR are listed here \cite{TV91, DACB96, RC01, YWB03, SW04, BYW04, DI06, SD09, Ada11, Li16}. In particular, Li developed the anisotropic mesh adaptation for image representation framework (AMA representation) in \cite{Li16} that provides better representation quality with reasonable computational cost than other MbIR methods.

Although CS is much more popular than MbIR, it is not clear if it performs better than MbIR in terms of reconstruction quality, especially for natural images that are not sparse on their own. This paper aims to perform a preliminary comparison between CS and MbIR for digital images from different sources and provide guidance for further research efforts in the area of image representation.

The rest of the paper is organized as follows. Section \ref{sec-rep} gives a brief description of compressive sampling (CS) as well as mesh-based image representation. Some CS algorithms and a recently developed MbIR method, anisotropic mesh adaptation for image representation (AMA representation), are introduced. Section \ref{sec-comparison} describes the comparison metrics and presents the results for images of different styles and resolutions. A short conclusion is given in Section \ref{sec-conclusion}.

\section{Compressive sampling and mesh-based image representation}
\label{sec-rep}
In this section, we briefly introduce the two image representation approaches: compressive sampling and adaptive sampling, in particular, anisotropic mesh adaptation for image representation (AMA representation) that is a recently developed anisotropic mesh-based image representation framework.

\subsection{Compressive sampling}
\label{sec-cs}
The key idea of compressive sampling (CS) is the sparsity of the signal that needs to be reconstructed. A signal is called sparse if most of its components are zero. It is possible to represent a sparse signal using a small number of measurements and reconstruct it by solving an underdetermined linear system. By reducing the sampling rate, CS can improve the signal acquisition and transmission speed and reduce hardware requirements for sensors. It takes more computational effort during reconstruction on the receiver side.

Let $\vec{x} \in \mathbb{C}^N$ be a sparse vector needs to be reconstructed from the underdetermined measurements $\vec{x}_{meas} = A \vec{x} \in \mathbb{C}^m$ with $m<<N$. Then there are two key steps in CS.
\begin{enumerate}
  \item Design the measurement (also called sensing) matrix $A \in \mathbb{C}^{m \times N}$ so that as much intrinsic information will contribute to the reconstruction.
  \item Solve $\vec{x}$ from the underdetermined linear system $\vec{x}_{meas} = A \vec{x}$ with balance of efficiency and accuracy.
\end{enumerate}

For image representation purpose, we convert the 2D images into a vector $\vec{x} \in \mathbb{R}^N$, which serves as the ground truth. Although $\vec{x}$ may not be sparse on its own, it can be sparse in wavelet or Fourier domain. Thus, $m (<<N)$ sample points are randomly chosen from $\vec{x}$ in wavelet or Fourier domain. Then a random matrix $A$ of the corresponding size is generated according to the sample points. The measurement data is simulated by multiplying $A$ and $\vec{x}$ as $\vec{x}_{meas} = A \vec{x} \in \mathbb{R}^m$. The ratio $m/N$ is denoted as the \textit{sample density} of the representation.

For the reconstruction step, we consider two methods here. One is to solve the $l_1$-minimization problem by using the log-barrier algorithm, and the $l_1$-MAGIC package \cite{L1magic} (written in MATLAB) is used for the computations. The other is the iterative soft thresholding algorithm (ISTA), and the \textit{pit} package \cite{pit} (written in Python) is used for the computations. Minor modifications of the program codes have been made to display the reconstructed image as well as compute the quality of reconstruction.

\subsection{AMA representation}
\label{sec-amir}
Different from CS, adaptive sampling works directly on image pixels. The idea is to choose the essential pixels of the image and discard non-important ones. Various approaches, especially mesh-based image representation methods, have been developed since the 1990s. Here, we briefly introduce the recently developed AMA representation framework \cite{Li16} that consists of four key steps.
\begin{enumerate}
  \item Generate an initial triangular mesh based on the desired sample density. The sample density determines the number of vertices in the mesh. Delaunay triangulation can be used to generate the initial mesh.
  \item Assign values to mesh vertices from original image grey values using linear finite element interpolation and compute the user-defined metric tensor $\mathbb{M}$ on the mesh.
  \item Adapt the mesh to be a quasi-$\mathbb{M}$-uniform mesh that almost fits the provided metric tensor $\mathbb{M}$.
  \item Reconstruct the image using the final adaptive mesh with finite element interpolation for triangles.
\end{enumerate}
Steps 2 and 3 can be repeated a few times in order to obtain a good adaptive mesh, which is the part that takes most computational resources in this framework.

Li also proposed the GPRAMA method in \cite{Li16} that combines the greedy-point-removal algorithm with AMA representation using mesh patching techniques. GPRAMA can provide better representation quality than AMA but is computationally more expensive due to the greedy-point-removal procedure. As a preliminary investigation, we only consider AMA representation in this paper.

\section{Comparison results}
\label{sec-comparison}
In this section, we present the results obtained for eight images of different styles and resolutions. The parameter values for CS algorithms are taken from the packages as default. For AMA representation, the anisotropic metric tensor $\mathbb{M}_{aniso}$ (Eq. (8) in \cite{Li16}) is used.

In $l1$-MAGIC package, the \textit{CS\_tveq()} function solves a total variation minimization problem with equality constraints. For the tested images, this function provides better reconstruction quality as well as takes less time than \textit{CS\_tvqc()} function that solves a total variation minimization problem with quadratic constraints. Although \textit{CS\_tvqc} with quantization runs faster than \textit{CS\_tveq()}, the reconstruction quality is worse. Thus, only the results obtained from \textit{CS\_tveq()} function are presented in this paper and denoted as TVeq.

The results obtained from the Python package \textit{pit} using the iterative soft thresholding algorithm are denoted as ISTA. And the results obtained from AMA representation are denoted as AMA.

\subsection{Comparison metrics}
\label{sec-metrics}

To compare the quality of the reconstruction from different approaches, we consider two commonly used metrics: peak-signal-to-noise-ratio (PSNR) and structural similarity index measure (SSIM).

PSNR estimates the absolute error between the reconstructed image and the original image. It is calculated based on the mean square error (MSE) as follows.
\begin{equation}
  \label{psnr}
  \text{PSNR} = 20\log_{10} \left(\frac{2^p - 1}{\sqrt{MSE}} \right),
\end{equation}
where $p$ is the sample precision. For grayscale images, $p=8$. The \textit{psnr()} function in MATLAB is used for the calculation. Larger PSNR value indicates better representation quality.

On the other hand, SSIM considers the inter-dependence of the pixels that carries important information about the structure of the objects in the image. The measure between two images $x$ and $y$ of common size is calculated as follows
\begin{equation}
  \label{ssim}
  \text{SSIM}(x, y) = \frac{(2\mu_x \mu_y + C_1)(2\sigma_{xy} + C_2)}
  {(\mu_x^2+\mu_y^2+C_1)(\sigma_x^2+\sigma_y^2+C_2)},
\end{equation}
where $\mu_{i}$ and $\sigma_{i}$ are average and variance of $i$ ($i=x, y$), respectively, and $\sigma_{xy}$ is the covariance of $x$ and $y$. The \textit{ssim()} function in MATLAB is used for the calculation that also provides the local SSIM map. Larger SSIM value indicates better representation quality.

\subsection{Comparison results}
\label{sec-results}

\begin{figure}[ht!]
\centering
\hspace{2mm}
\hbox{
\begin{minipage}[b]{1.5in}
\includegraphics[width=1.5in]{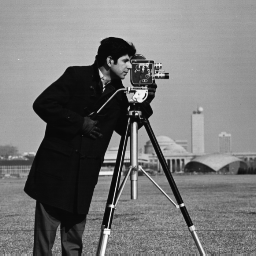}
\centerline{(1)}
\end{minipage}
\hspace{2mm}
\begin{minipage}[b]{1.5in}
\includegraphics[width=1.5in]{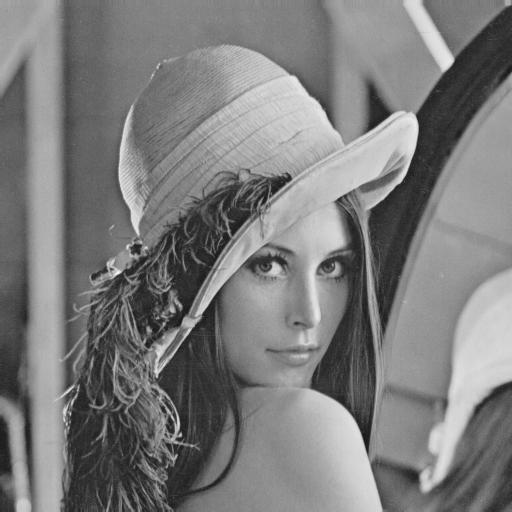}
\centerline{(2)}
\end{minipage}
\hspace{2mm}
\begin{minipage}[b]{1.5in}
\includegraphics[width=1.5in]{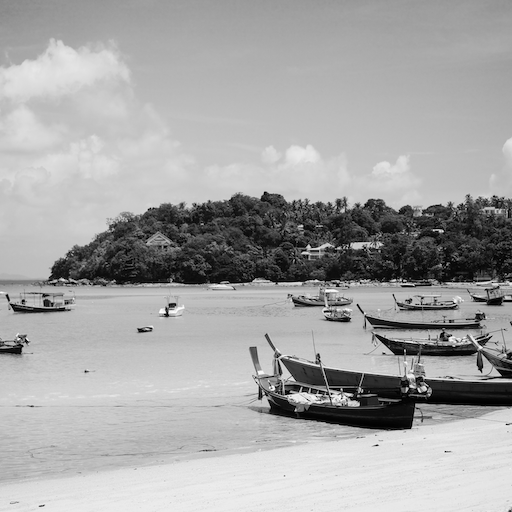}
\centerline{(3)}
\end{minipage}
\hspace{2mm}
\begin{minipage}[b]{1.5in}
\includegraphics[width=1.5in]{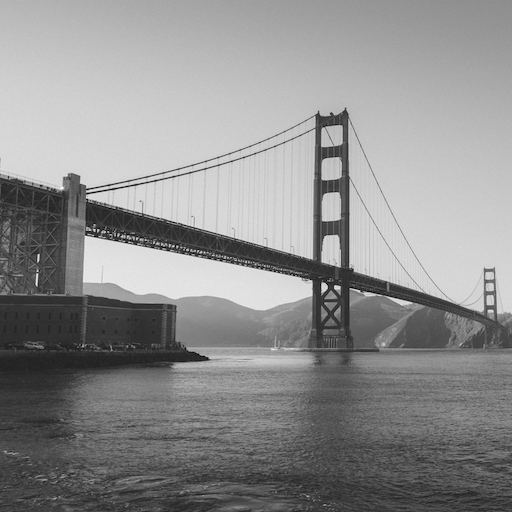}
\centerline{(4)}
\end{minipage}
}
\vspace{5mm}
\hbox{
\begin{minipage}[b]{1.5in}
\includegraphics[width=1.5in]{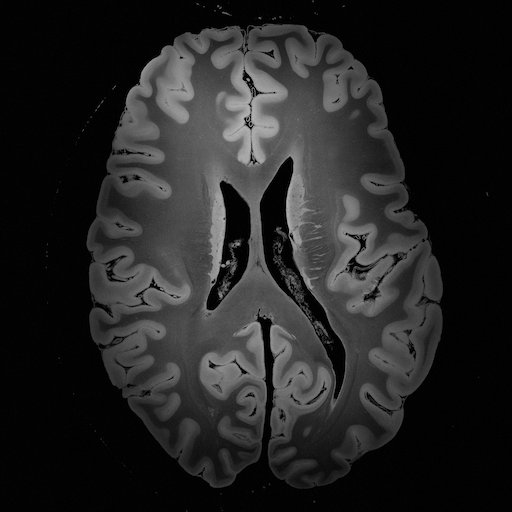}
\centerline{(5)}
\end{minipage}
\hspace{2mm}
\begin{minipage}[b]{1.5in}
\includegraphics[width=1.5in]{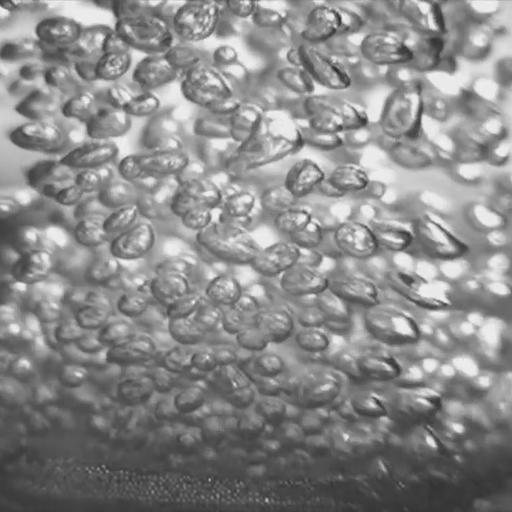}
\centerline{(6)}
\end{minipage}
\hspace{2mm}
\begin{minipage}[b]{1.5in}
\includegraphics[width=1.5in]{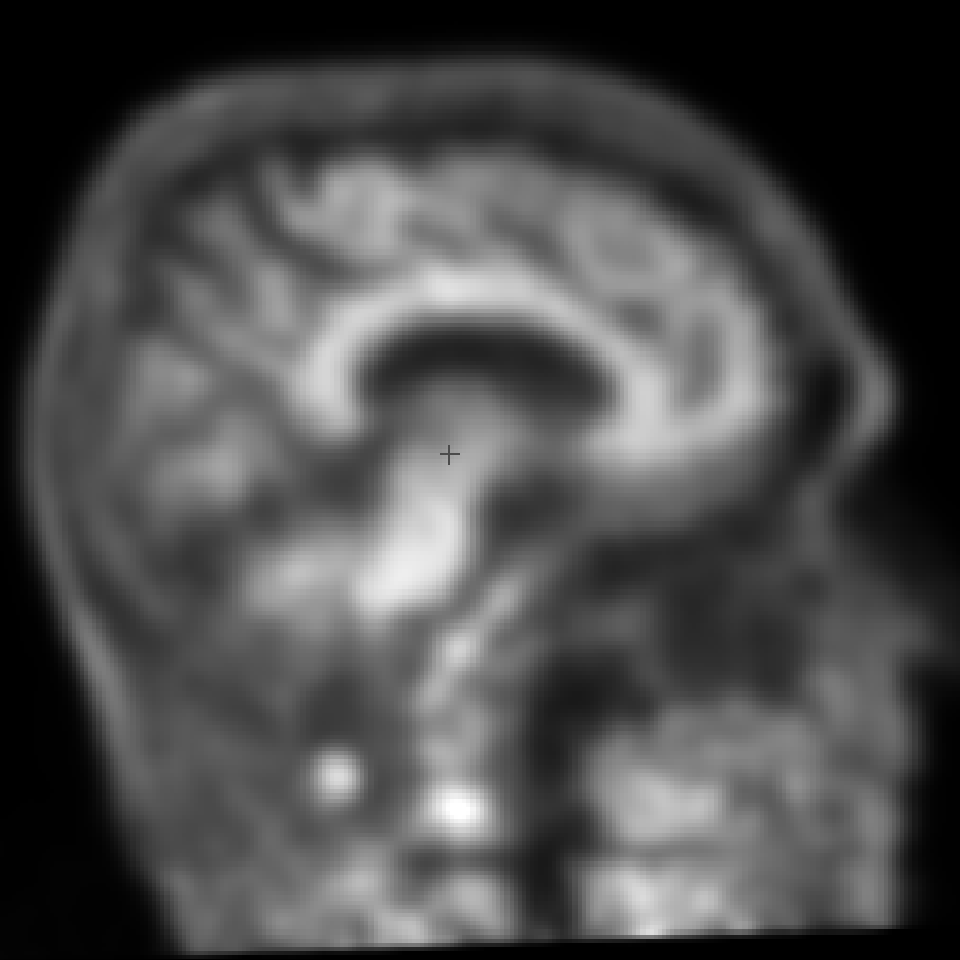}
\centerline{(7)}
\end{minipage}
\hspace{2mm}
\begin{minipage}[b]{1.5in}
\includegraphics[width=1.5in]{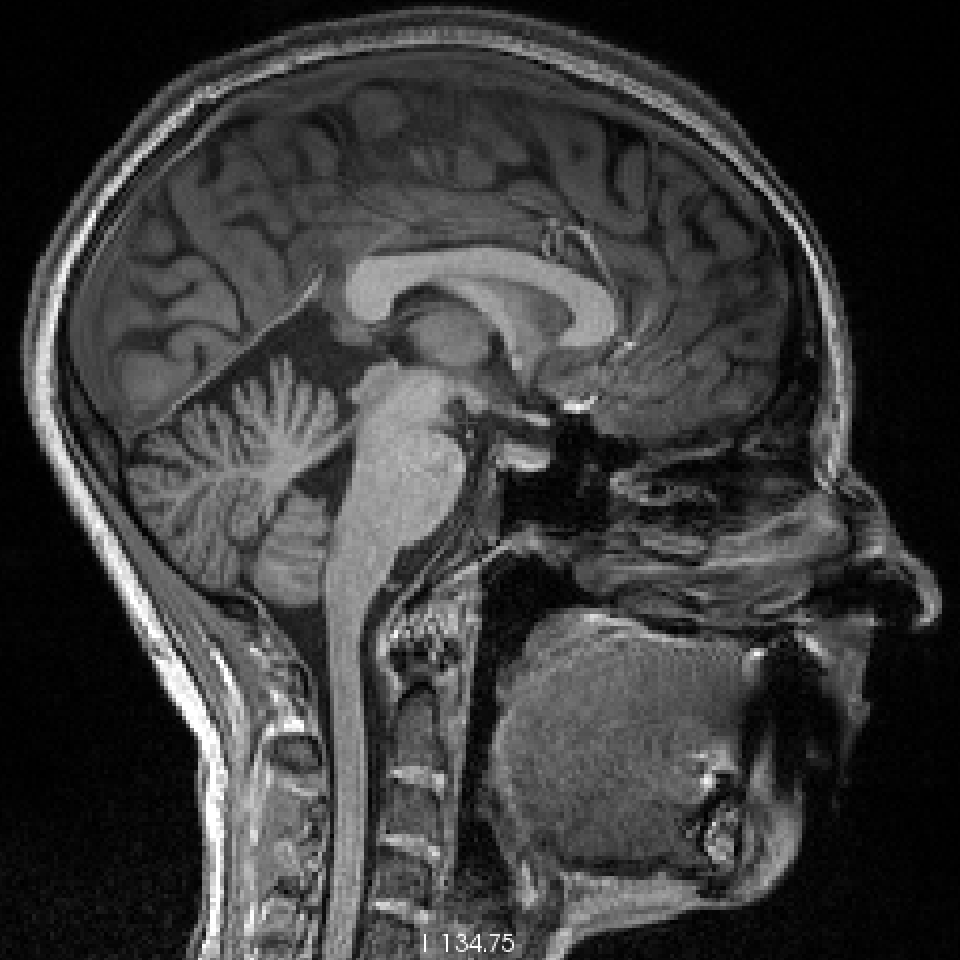}
\centerline{(8)}
\end{minipage}
}
\caption{Original images: (1) cameraman, $256 \times 256$, (2) Lena, $512 \times 512$, (3) beach, $730 \times 730$, (4) bridge, $3000 \times 3000$, (5) brain MRI, $1760 \times 1760$, (6) heat bubble, $720 \times 720$, (7) AD (PET), $960 \times 960$, and (8)migraine (MRI), $960 \times 960$.}
\label{fig-imgs}
\end{figure}

Fig. \ref{fig-imgs} shows eight images chosen for this comparison task, all of which are converted into grayscale. The first four images are publicly available from USC-SIPI or MATLAB image folder. The high-resolution brain MR image Fig. \ref{fig-imgs}(5) is one of the slices extracted from a 7-Tesla MRI of the ex vivo human brain at 100-micron resolution by Edlow \textit{et al} \cite{EMH19}. The last three images are provided by the ASU-Mayo Center for Innovative Imaging. 

The results for sample density at 3\% are shown in Table \ref{table-1}, while those for sample density at 10\% are shown in Table \ref{table-2}. It is clear that the representation quality improves when sample density increases because more information is embedded in the measurement data.

For results in each table, that is, at the same sample density, AMA provides the best representation quality in terms of both PSNR and SSIM. ISTA performs the worst in terms of the reconstruction quality, which might be due to the default threshold values used in the code. Finding optimal values for the parameters could improve the performance. However, this comparison results deliver an overall message that AMA representation can provide better quality than CS algorithms.

\begin{table}[thb!]
\caption{Comparison of representation qualities at sample density 3\%}
\vspace{2pt}
\centering
\begin{tabular}{|c|c||c|c|c||c|c|c|}
\hline
 image & resolution & \multicolumn{3}{c||}{PSNR (dB)} & \multicolumn{3}{c|}{SSIM} \\ \cline{3-8}
 & & TVeq & ISTA & AMA & TVeq & ISTA & AMA \\
\hline
 cameraman & $256 \times 256$ & 30.69 & 18.90 & 33.09 & 0.61 & 0.54 & 0.74 \\
 \hline
 Lena & $512 \times 512$ & 31.98 & 23.59 & 35.23 & 0.74 & 0.69 & 0.85 \\
 \hline
 beach & $730 \times 730$ & 32.82 & 21.35 & 35.04 & 0.75 & 0.69 & 0.83 \\
 \hline
 bridge & $3000 \times 3000$ & 34.91 & 27.31 & 36.92 & 0.80 & 0.74 & 0.85  \\
 \hline
 brain MRI & $1760 \times 1760$ & 35.38 & 28.24 & 36.15 & 0.68 & 0.65 & 0.71 \\
 \hline
 heat bubble$^*$ & $720 \times 720$ & 30.80 & 25.23 & 34.91 & 0.73 & 0.77 & 0.87 \\
 \hline
 AD (PET)$^*$ & $960 \times 960$ & 40.82 & 38.53 & 44.49 & 0.94 & 0.95 & 0.96 \\
 \hline
 migraine (MRI)$^*$ & $960 \times 960$ & 32.08 & 24.27 & 34.09 & 0.69 & 0.68 & 0.78 \\
\hline
\end{tabular} \\
($^*$: provided by ASU-Mayo Center for Innovative Imaging.)
\vspace{2pt}
\label{table-1}
\end{table}

\begin{table}[thb!]
\caption{Comparison of representation qualities at sample density 10\%}
\vspace{2pt}
\centering
\begin{tabular}{|c|c||c|c|c||c|c|c|}
\hline
 image & resolution & \multicolumn{3}{c||}{PSNR (dB)} & \multicolumn{3}{c|}{SSIM} \\ \cline{3-8}
 & & TVeq & ISTA & AMA & TVeq & ISTA & AMA \\
\hline
 cameraman & $256 \times 256$ & 32.29 & 21.21 & 34.44 & 0.73 & 0.66 & 0.82 \\
 \hline
 Lena & $512 \times 512$ & 34.51 & 27.26 & 38.22 & 0.84 & 0.80 & 0.92 \\
 \hline
 beach & $730 \times 730$ & 34.28 & 23.68 & 36.59 & 0.81 & 0.78 & 0.90 \\
 \hline
 bridge & $3000 \times 3000$ & 37.58 & 30.86 & 40.50 & 0.87 & 0.82 & 0.92  \\
 \hline
 brain MRI & $1760 \times 1760$ & 36.09 & 30.05 & 36.84 & 0.72 & 0.70 & 0.74 \\
 \hline
 heat bubble$^*$ & $720 \times 720$ & 34.43 & 31.34 & 40.94 & 0.87 & 0.89 & 0.96 \\
 \hline
 AD (PET)$^*$ & $960 \times 960$ & 47.59 & 40.23 & 47.36 & 0.98 & 0.96 & 0.98 \\
 \hline
 migraine (MRI)$^*$ & $960 \times 960$ & 35.41 & 26.94 & 36.21 & 0.84 & 0.77 & 0.87 \\
\hline
\end{tabular} \\
($^*$: provided by ASU-Mayo Center for Innovative Imaging.)
\vspace{2pt}
\label{table-2}
\end{table}

It is interesting to observe that, for the high-resolution brain MR image in Fig. \ref{fig-imgs}(5), TVeq performance is close to that of AMA in terms of both PSNR and SSIM. At the same time, ISTA is also close in terms of SSIM. Similar results are observed for AD (PET) image in Fig. \ref{fig-imgs}(7). The reason might be due to the high sparsity of those two images.

\begin{figure}[ht!]
\centering
\hspace{2mm}
\hbox{
\begin{minipage}[b]{1.5in}
\includegraphics[width=1.5in]{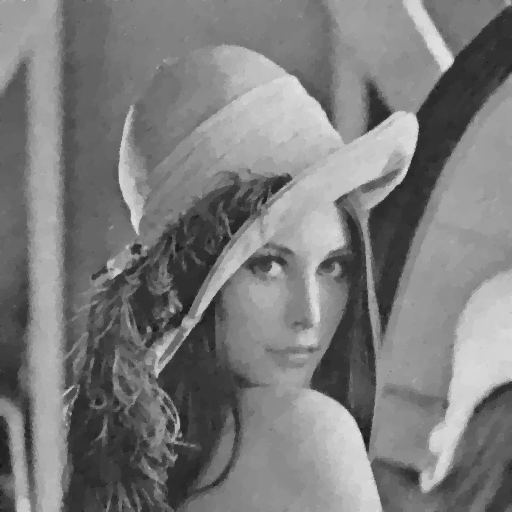}
\centerline{(a)}
\end{minipage}
\hspace{2mm}
\begin{minipage}[b]{1.5in}
\includegraphics[width=1.5in]{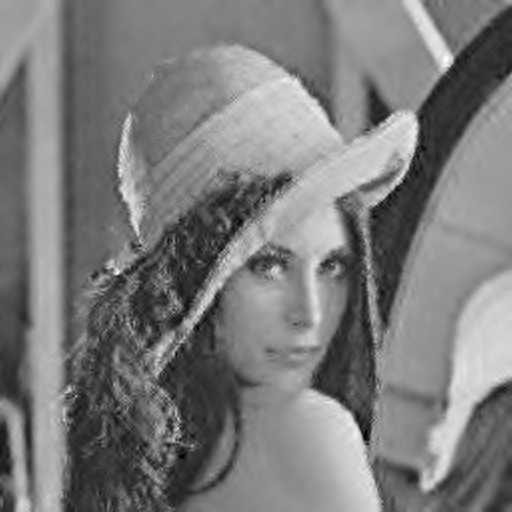}
\centerline{(b)}
\end{minipage}
\hspace{2mm}
\begin{minipage}[b]{1.5in}
\includegraphics[width=1.5in]{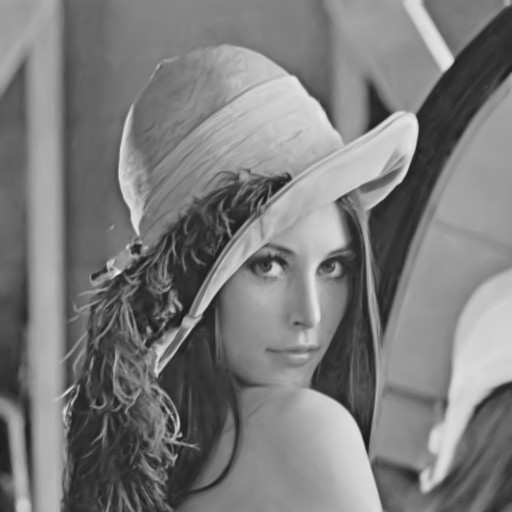}
\centerline{(c)}
\end{minipage}
}
\vspace{5mm}
\hspace{2mm}
\hbox{
\begin{minipage}[b]{1.5in}
\includegraphics[width=1.5in]{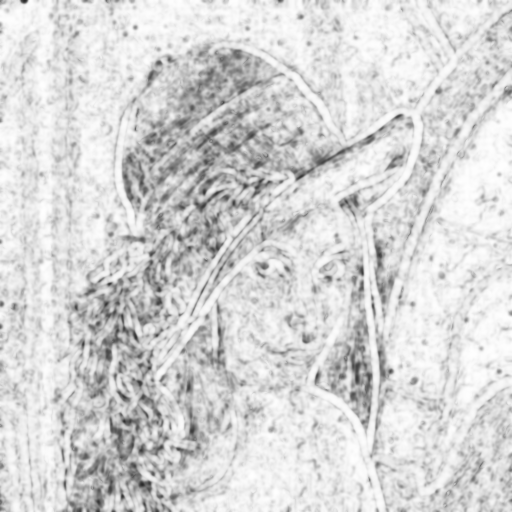}
\centerline{(d)}
\end{minipage}
\hspace{2mm}
\begin{minipage}[b]{1.5in}
\includegraphics[width=1.5in]{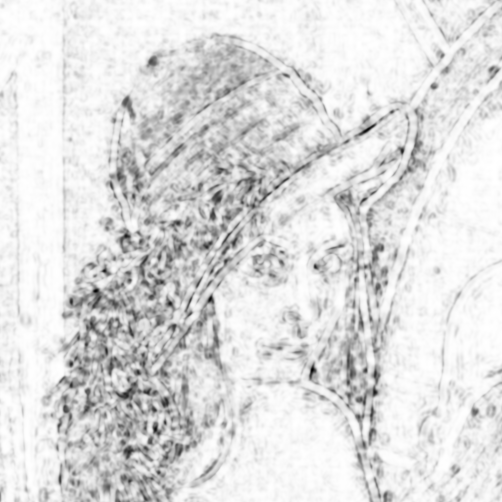}
\centerline{(e)}
\end{minipage}
\hspace{2mm}
\begin{minipage}[b]{1.5in}
\includegraphics[width=1.5in]{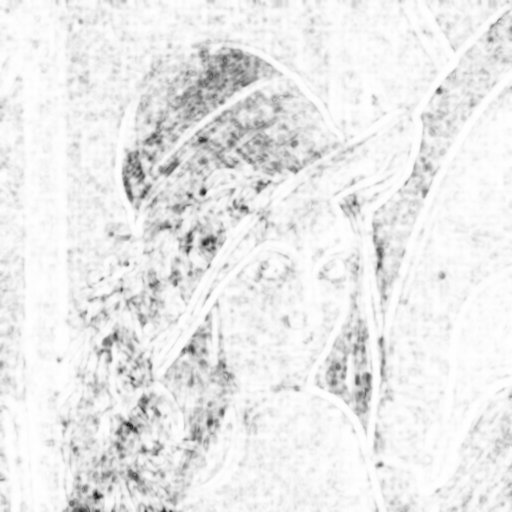}
\centerline{(f)}
\end{minipage}
}
\caption{Reconstruction of image Lena at sample density 10\%: (a) TVeq, PSNR=34.51, (b) ISTA, PSNR=27.26, (c) AMA, PSNR=38.22, (d) TVeq SSIM map, SSIM=0.84, (e) ISTA SSIM map, SSIM=0.80, and (f) AMA SSIM map, SSIM=0.92.}
\label{fig-results-lena}
\end{figure}

\begin{figure}[ht!]
\centering
\hspace{2mm}
\hbox{
\begin{minipage}[b]{1.5in}
\includegraphics[width=1.5in]{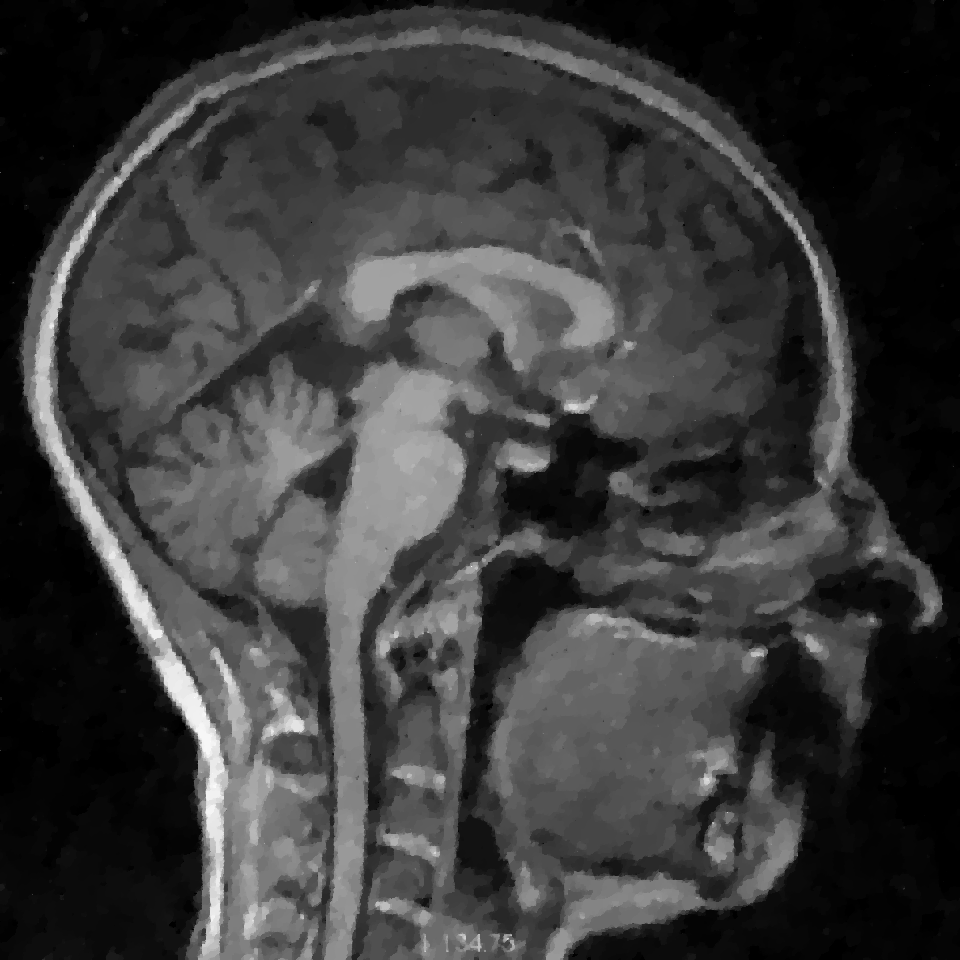}
\centerline{(a)}
\end{minipage}
\hspace{2mm}
\begin{minipage}[b]{1.5in}
\includegraphics[width=1.5in]{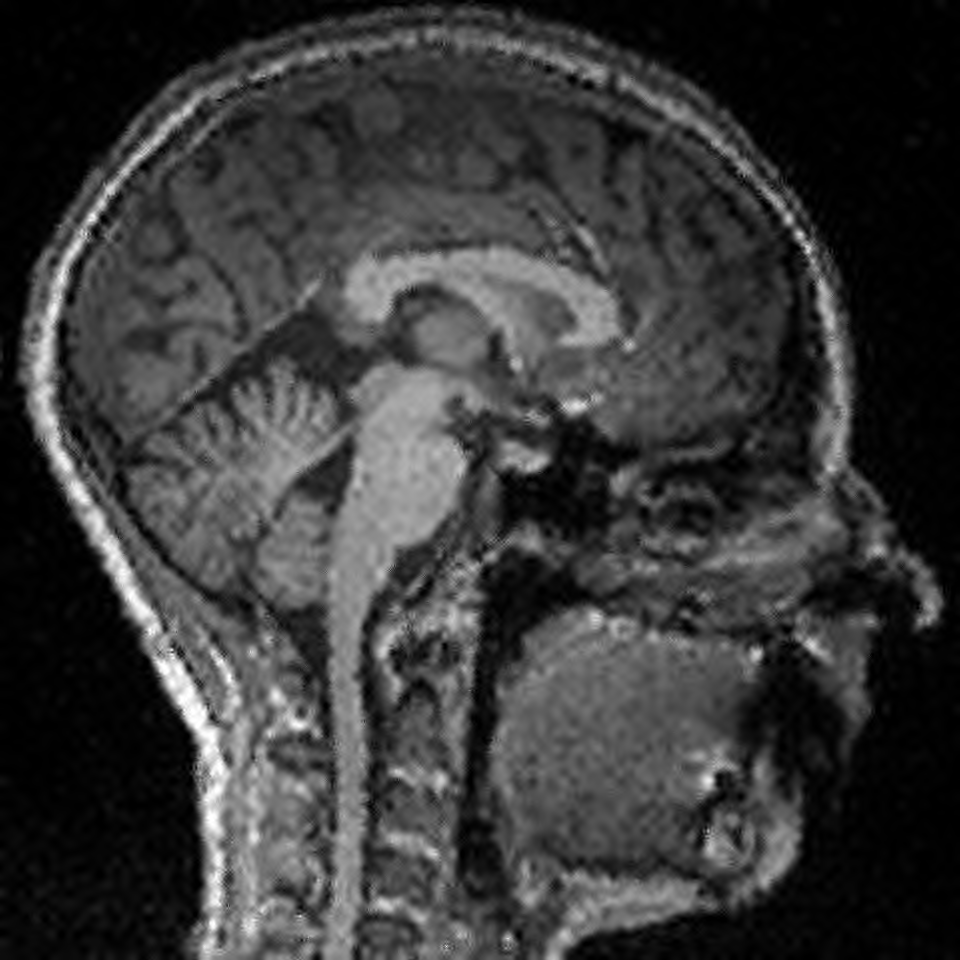}
\centerline{(b)}
\end{minipage}
\hspace{2mm}
\begin{minipage}[b]{1.5in}
\includegraphics[width=1.5in]{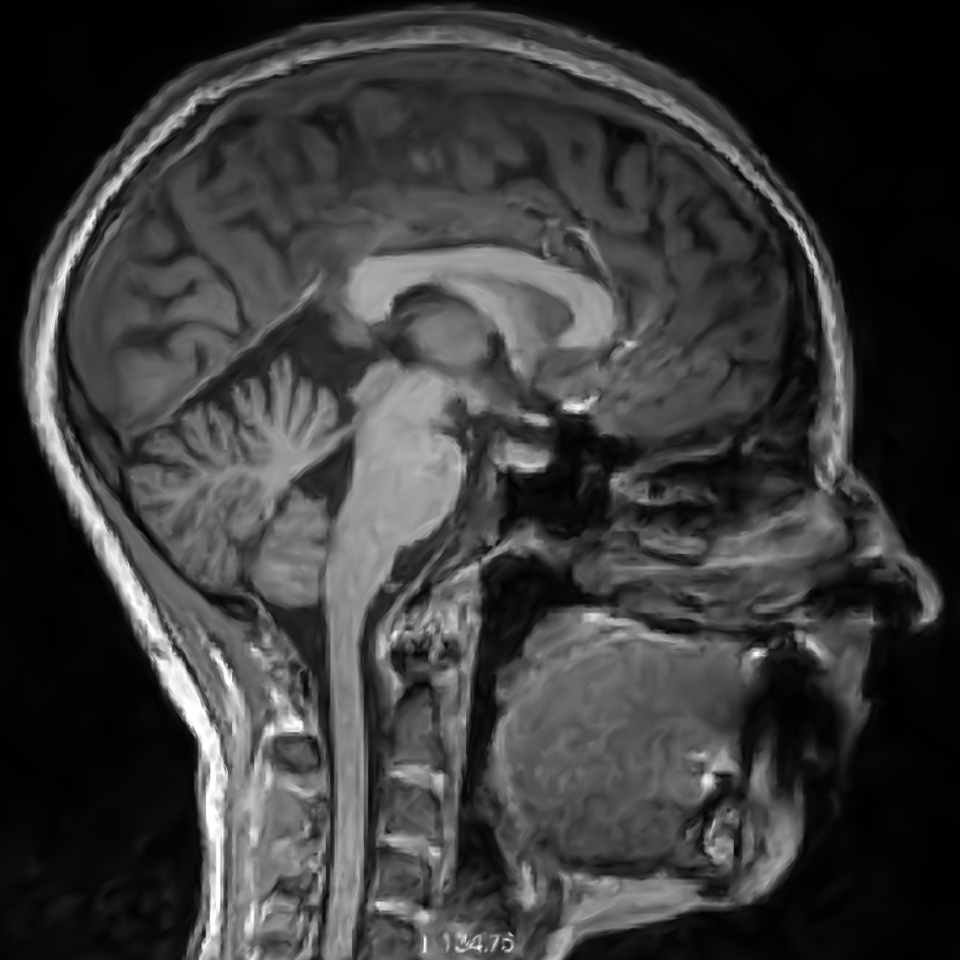}
\centerline{(c)}
\end{minipage}
}
\vspace{5mm}
\hspace{2mm}
\hbox{
\begin{minipage}[b]{1.5in}
\includegraphics[width=1.5in]{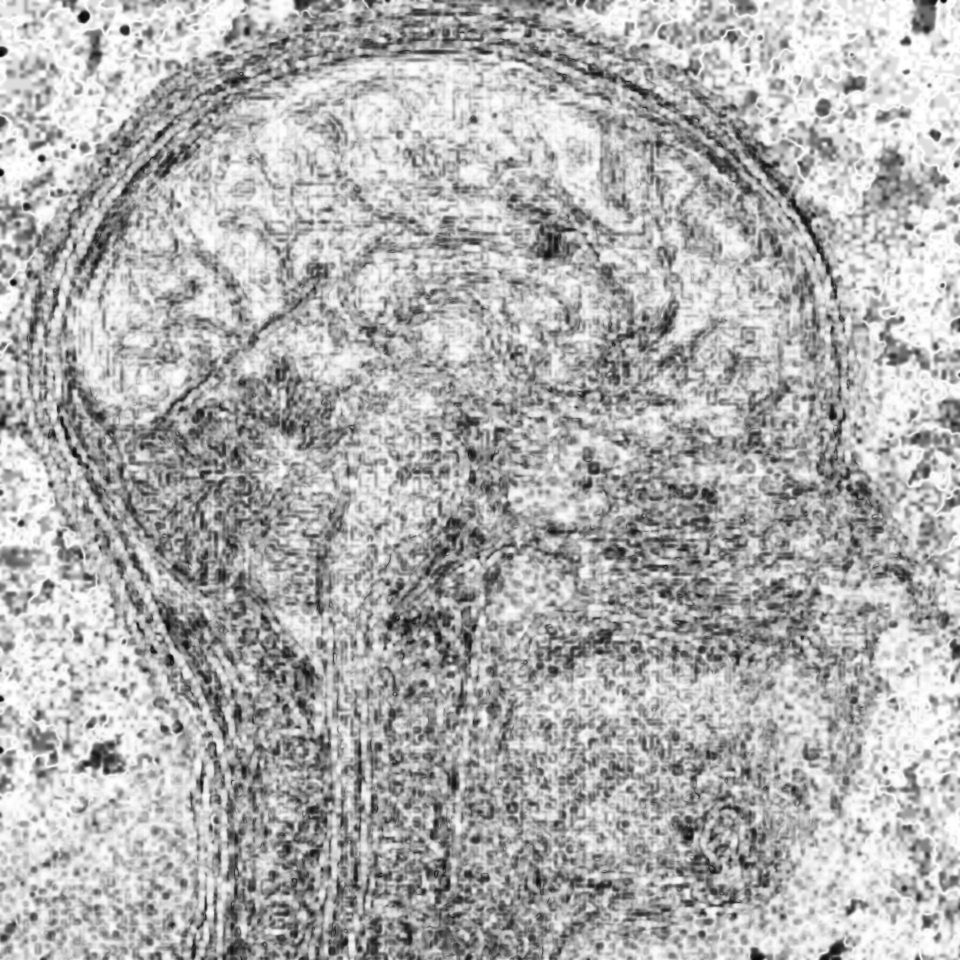}
\centerline{(d)}
\end{minipage}
\hspace{2mm}
\begin{minipage}[b]{1.5in}
\includegraphics[width=1.5in]{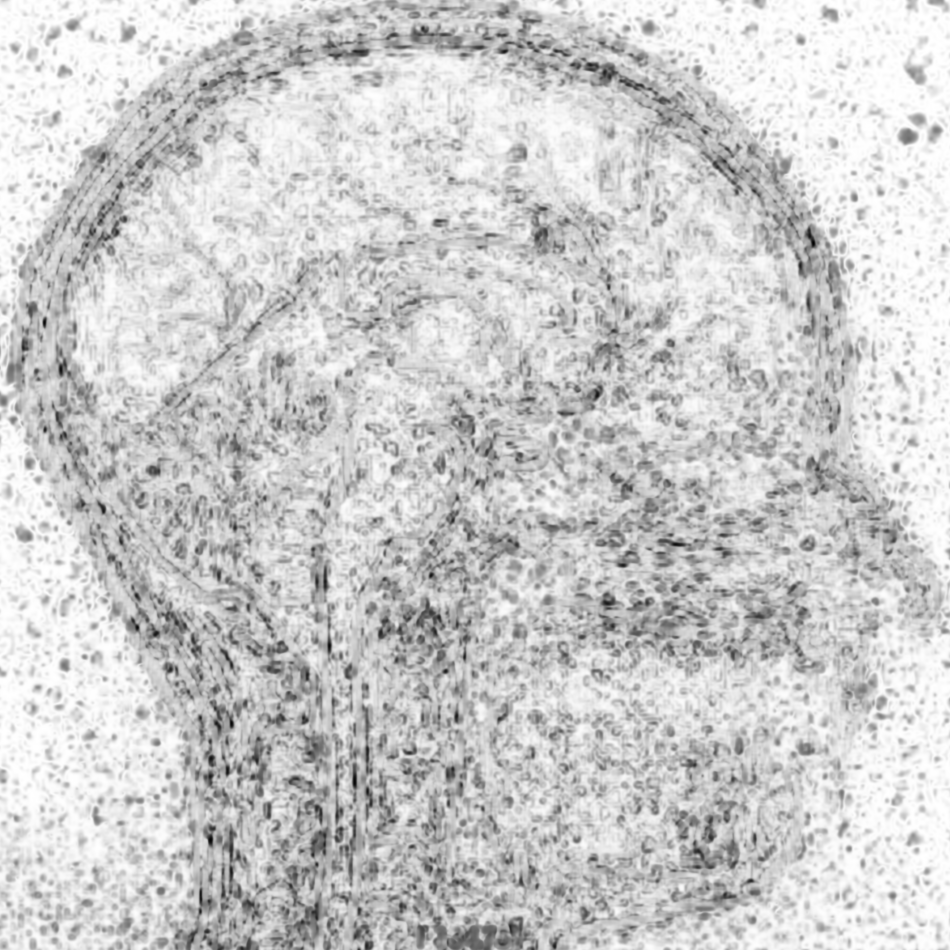}
\centerline{(e)}
\end{minipage}
\hspace{2mm}
\begin{minipage}[b]{1.5in}
\includegraphics[width=1.5in]{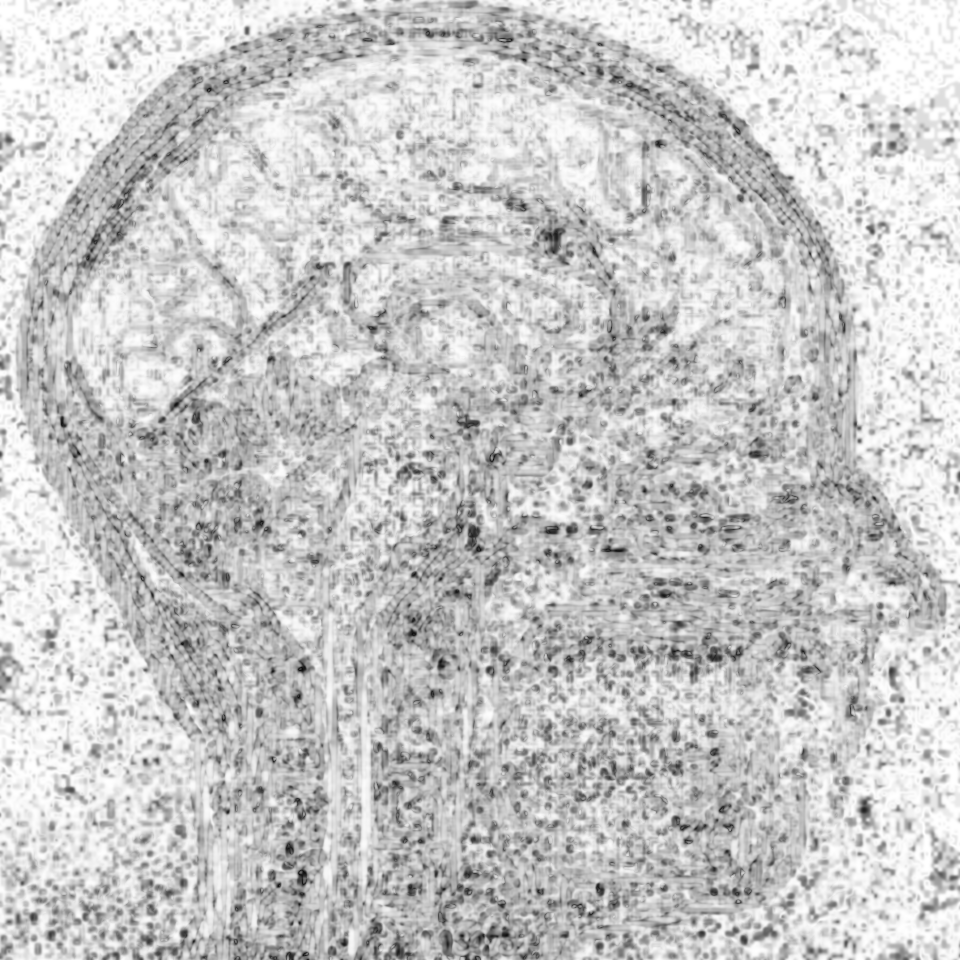}
\centerline{(f)}
\end{minipage}
}
\caption{Reconstruction of image migraine (MRI) at sample density 3\%: (a) TVeq, PSNR=32.08, (b) ISTA, PSNR=24.27, (c) AMA, PSNR=34.09, (d) TVeq SSIM map, SSIM=0.69, (e) ISTA SSIM map, SSIM=0.68, and (f) AMA SSIM map, SSIM=0.78.}
\label{fig-results-migraine}
\end{figure}

Fig. \ref{fig-results-lena} shows the reconstruction results for image Lena at sample density 10\%, and Fig. \ref{fig-results-migraine} shows the reconstruction results for image migraine (MRI) at sample density 3\%. The results for other images are similar.

\begin{figure}[ht!]
\centering
\hspace{2mm}
\hbox{
\begin{minipage}[b]{1.5in}
\includegraphics[width=1.5in]{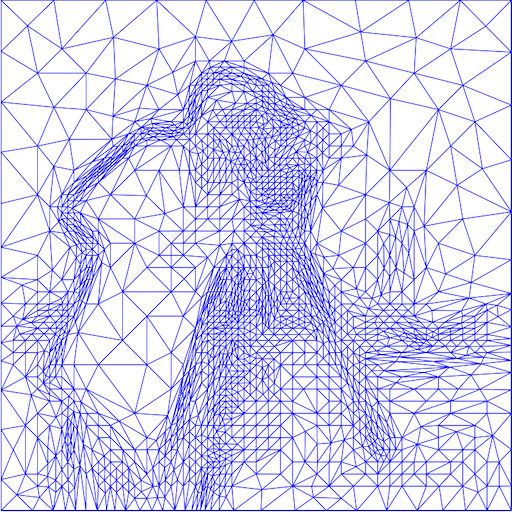}
\centerline{(1)}
\end{minipage}
\hspace{2mm}
\begin{minipage}[b]{1.5in}
\includegraphics[width=1.5in]{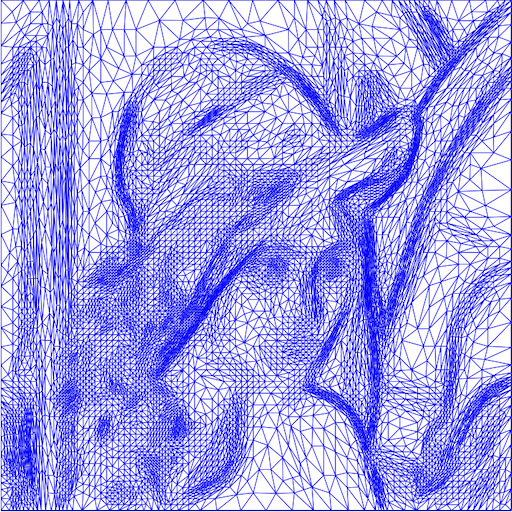}
\centerline{(2)}
\end{minipage}
\hspace{2mm}
\begin{minipage}[b]{1.5in}
\includegraphics[width=1.5in]{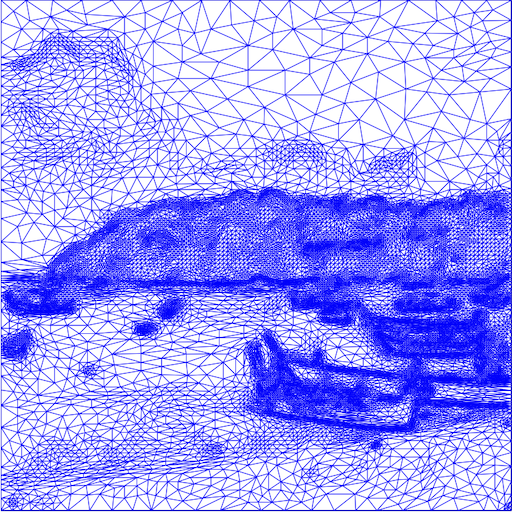}
\centerline{(3)}
\end{minipage}
\hspace{2mm}
\begin{minipage}[b]{1.5in}
\includegraphics[width=1.5in]{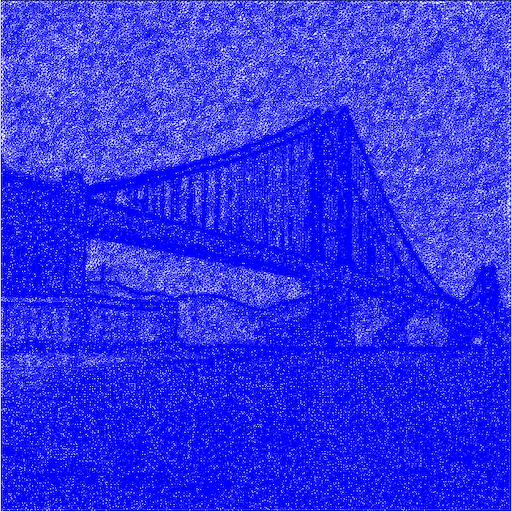}
\centerline{(4)}
\end{minipage}
}
\vspace{5mm}
\hbox{
\begin{minipage}[b]{1.5in}
\includegraphics[width=1.5in]{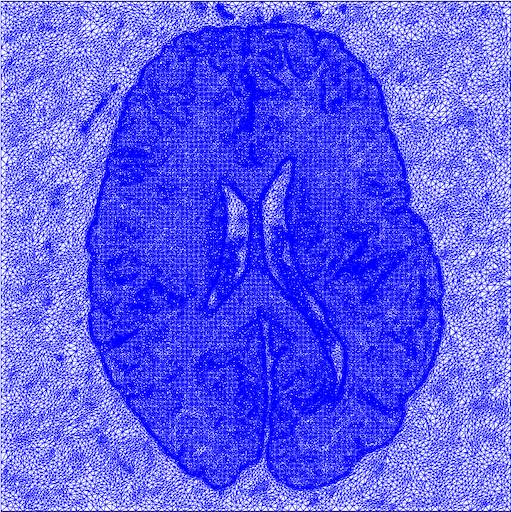}
\centerline{(5)}
\end{minipage}
\hspace{2mm}
\begin{minipage}[b]{1.5in}
\includegraphics[width=1.5in]{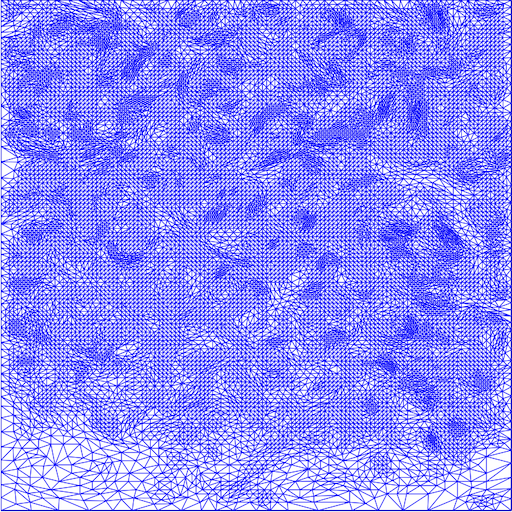}
\centerline{(6)}
\end{minipage}
\hspace{2mm}
\begin{minipage}[b]{1.5in}
\includegraphics[width=1.5in]{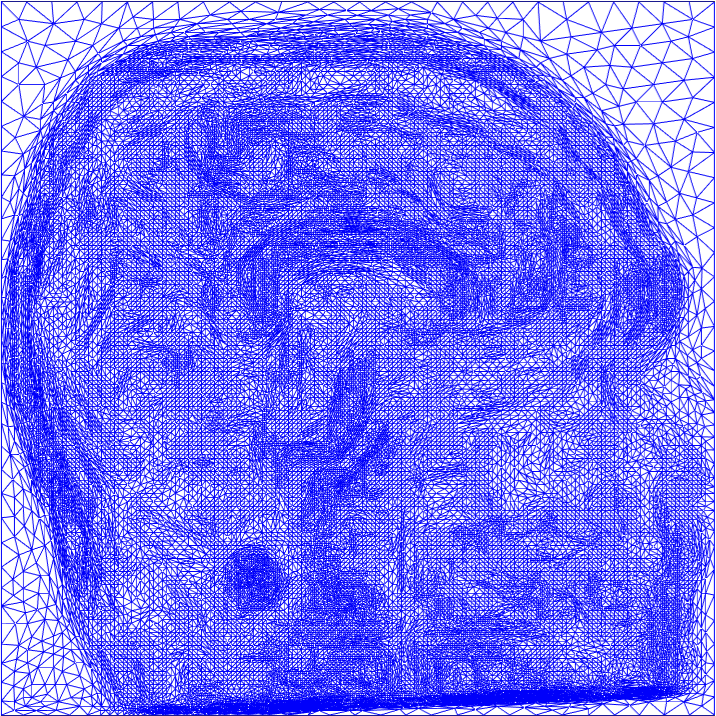}
\centerline{(7)}
\end{minipage}
\hspace{2mm}
\begin{minipage}[b]{1.5in}
\includegraphics[width=1.5in]{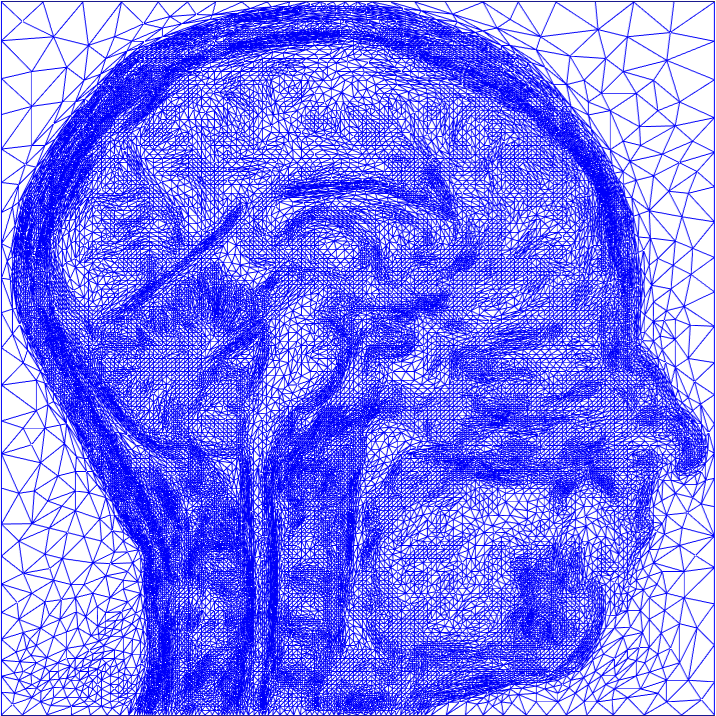}
\centerline{(8)}
\end{minipage}
}
\caption{Anisotropic triangular meshes for AMA representation at sample density \%3: (1) cameraman, (2) Lena, (3) beach, (4) bridge, (5) brain MRI, (6) heat bubble, (7) AD (PET), and (8) migraine (MRI).}
\label{fig-meshes}
\end{figure}

Fig. \ref{fig-meshes} shows the anisotropic triangular meshes for the original images produced by AMA representation at sample density 3\%. As can be seen, the triangular elements concentrate around the edges of the images so that they can represent the key features of the images.

\section{Conclusions}
\label{sec-conclusion}

In this paper, we have performed a preliminary comparison between compressive sampling and adaptive sampling, specifically, AMA representation. Due to the large number of methods developed in the CS field, this comparison is by no means thorough. However, the work does provide some guidance for future efforts in the area of image representation.

Our results indicate that AMA representation can provide better representation quality than CS algorithms. CS works well for ideal sparse signals but may not for natural images. Although different measurement matrix can be designed, and various parameter values in different algorithms can be optimized, the improvement of CS performance, especially on natural images, may still be modest. Further comparison with recently developed CS algorithms is needed.

On the other hand, AMA representation works directly on image pixels by performing anisotropic mesh adaptation to concentrate triangular elements around the edges of the images. Although some parameters are used for the mesh adaptation procedure, their default values in the program code work well for general problems. The representation procedure is straightforward without the need to specify different parameters other than the desired sample density. Note that the AMA representation quality can still be improved by adopting the greedy-point-removal algorithm (GPRAMA) at the expense of more computational cost.

Besides better representation quality than compressive sampling, AMA representation can also be applied to improve the efficiency of partial differential equation based image processing such as image segmentation.

\end{document}